\documentclass[12pt, twocolumn,showpacs,preprintnumbers,aps,prl,reprint,superscriptaddress]{revtex4-1}

\usepackage{graphicx}
\usepackage{subfigure}
\usepackage{color}
\usepackage{amsmath}
\usepackage[linktocpage,colorlinks=true,linkcolor=blue,citecolor=blue,breaklinks=true]{hyperref}
\usepackage{verbatim}
\usepackage{breakcites}
\usepackage{pgfplots}

\begin{document}

\preprint{}

\title{Roughness as a Route to the Ultimate Regime of Thermal Convection}

\author{Srikanth Toppaladoddi}
\affiliation{Yale University, New Haven, Connecticut, USA}

\author{Sauro Succi}
\affiliation{ Istituto per le Applicazioni del Calcolo ``Mauro Picone" (C.N.R.), Rome, Italy}

\author{John S. Wettlaufer}
\affiliation{Yale University, New Haven, Connecticut, USA}
\affiliation{Mathematical Institute, University of Oxford, Oxford, UK}
\affiliation{Nordita, Royal Institute of Technology and Stockholm University, Stockholm, Sweden}

\email[]{john.wettlaufer@yale.edu}

\date{\today}

\begin{abstract}
{We use highly resolved numerical simulations to study turbulent Rayleigh-B\'enard convection in a cell with sinusoidally rough upper and lower surfaces in two dimensions for $Pr = 1$ and $Ra = \left[4 \times 10^6, 3 \times 10^9\right]$. By varying the wavelength $\lambda$ at a fixed amplitude, we find an optimal wavelength  $\lambda_{\text{opt}}$ for which the Nusselt-Rayleigh scaling relation is $\left(Nu-1 \propto Ra^{0.483}\right)$ maximizing the heat flux.  This is consistent with the upper bound of Goluskin and Doering \cite{Goluskin:2016} who prove that $Nu$ can grow no faster than ${\cal O} (Ra^{1/2})$ as $Ra \rightarrow \infty$, and thus the concept that roughness facilitates the attainment of the so-called ultimate regime.  Our data nearly achieve the largest growth rate permitted by the bound.  When  $\lambda \ll \lambda_{\text{opt}}$ and $\lambda \gg \lambda_{\text{opt}}$, the planar case is recovered, demonstrating how controlling the wall geometry manipulates the interaction between the boundary layers and the core flow.  Finally, for each $Ra$ we choose the maximum $Nu$ among all $\lambda$, and thus optimizing over all $\lambda$, to find $Nu_{\text{opt}} - 1 = 0.01 \times Ra^{0.444}$.}
\end{abstract}


\maketitle

The ubiquity and importance of thermal convection in many natural and man-made settings is well known \cite{kadanoff2001, Worster:2000, wettlaufer2011}. The simplest scenario that has been used to study the fundamental aspects of thermal convection is the Rayleigh-B\'enard system \cite{chandra2013}. The flow in this system is governed by three non-dimensional parameters: (1) the Rayleigh number 
$Ra = g \alpha \Delta T H^3/ \nu \kappa$, which is the ratio of buoyancy to viscous forces, where $g$ is the acceleration due to gravity, $\alpha$ the thermal expansion coefficient of the fluid, $\Delta T$ the temperature difference across a layer of fluid of depth $H$, $\nu$ the kinematic viscosity (or momentum diffusivity) and $\kappa$ the thermal diffusivity; (2) the Prandtl number, $Pr = \nu/\kappa$; and (3) the aspect ratio of the cell, $\Gamma$, defined as the ratio of its width to height.

The primary aim of the corpus of studies of turbulent Rayleigh-B\'enard convection has been to determine the Nusselt number, $Nu$, defined as the ratio of total heat flux to conductive heat flux (Eq. \ref{eqn:Nu}), as a function of the three governing parameters, viz., $Nu = Nu(Ra, Pr, \Gamma)$. For $Ra \gg 1$ and fixed $Pr$ and $\Gamma$, this relation is usually sought in the form of a power law: $Nu = A(Pr, \Gamma) Ra^{\beta}$, where $\beta$ has a fundamental significance for the mechanisms underlying the transport of heat.   

The classical theory of Priestley \cite{priestley1954}, Malkus \cite{malkus1954} and Howard \cite{howard1966} is based on the argument that as $Ra \rightarrow \infty$ the dimensional heat flux should become independent of the depth of the cell, resulting in $\beta = 1/3$. A consequence of this scaling is that the conductive boundary layers (BLs) at the upper and lower surfaces, which are separated by a well mixed interior, do not interact.


However, Kraichnan \cite{kraichnan1962} reasoned that for extremely large $Ra$ the BLs undergo a transition leading to the generation of smaller scales near the boundaries that increase the system's efficiency in transporting the heat, predicting that $Nu \sim \left[Ra/\left(\ln Ra\right)^3\right]^{1/2}$.  In this,  ``Kraichnan-Spiegel'' or ``ultimate regime'' ($\beta = 1/2$), it is argued that the heat flux becomes independent of the molecular properties of the fluid \cite[e.g.,][]{spiegel1971, grossmann2000}.  Experimental \cite{sreenivasan2000, niemela2006, urban2011, urban2012} and numerical \cite{verzicco2003, stevens2011, gayen2013} studies have found $\beta \approx 1/3$.  Chavanne \emph{et al.} \cite{chavanne1997} and He \emph{et al.} \cite{he2012} have reported observing transitions to $\beta = 0.39$ and $\beta = 0.38$ in their respective experiments and these findings continue to stimulate discussion \cite{Skrbek:2015, He:2016}.  Motivated by studies of shear flow, Borue and Orszag \cite{Borue:1997} used pseudo-spectral methods at three resolutions (64$^3$, 128$^3$, 256$^3$ and hence values of $Ra$) to study ``homogeneous'' convection, in which the BL's are effectively removed.  Whilst the highest resolution was not numerically converged, the other two resolutions led to a range of $\beta = 0.40 \pm 0.05$.  This idea was later used in Lattice Boltzmann simulations for $Ra = \left[8.64 \times 10^5, 1.38 \times 10^7\right]$, to find $\beta = 0.51 \pm 0.06$ \cite{lohse2003}, ascribing this to the ultimate regime.  


Recently, Waleffe \emph{et al.} \cite{waleffephys2015} and Sondak \emph{et al.} \cite{sondak2015} numerically computed the steady solutions to the Oberbeck-Boussinesq equations for $Ra \le 10^9$ and $1 \le Pr \le 100$ in two dimensions. By fixing $Ra$ and $Pr$, steady solutions for different horizontal wavenumbers, $\alpha$, were computed.  The solution that maximized heat transport, $Nu \equiv Nu_{opt}$, was called optimal, for which $\alpha \equiv \alpha_{opt}$ and $Nu_{opt} - 1 = 0.115 \times Ra^{0.31}$, which is in agreement with experiments \cite{sreenivasan2000}. Although they found that $\beta$ was independent of $Pr$, the Prandtl number did have considerable effect on the geometry of the coherent structures that transported heat. For $Pr > 7$, the scaling for the optimal wavenumber was found to be $\alpha_{\text{opt}} = 0.257 \times Ra^{0.256}$. The horizontally averaged optimal temperature profiles had the following features: (a) The BLs were always unstably stratified. (b) The core region was either stably ($Pr \le 7$) or unstably ($Pr >7$) stratified. (c) The transition regions between the core and BLs were always stably stratified. Thus, with small departures, these profiles correspond to the marginally stable profile of Malkus \cite{malkus1954}, with $\beta = 1/3$.

An important aspect emerging from the study of planar Rayleigh-B\'enard convection in two dimensions for $Pr \ge 1$ is that the flow field \cite{schmalzl2004} and the $Nu$-$Ra$ scaling relations \cite{doering2009, waleffephys2015, sondak2015} are similar to those in three dimensions. Thus, this correspondence permits one to understand the processes driving the heat transport using well resolved two-dimensional simulations.

It is clear that the value $\beta$ takes in the limit $Ra \rightarrow \infty$ depends on the interaction between the BLs and the core flow. To understand the role of BLs in thermal convection, Shen \emph{et al.} \cite{shen1996} introduced rough upper and lower surfaces made of pyramidal elements in a cylindrical cell. They found that these elements enhanced the production of plumes, which were directly injected into the core flow, leading to an increase in $Nu$. The increase in $Nu$ was due to an increase in the pre-factor in the $Nu$--$Ra$ scaling relation. 
Whereas subsequent experiments found no effect of periodic roughness on $\beta$ \cite{du1998, du2000, ciliberto1999}, later studies confirmed that the changes in the flow field brought about by surface roughness do increase the value of $\beta$ from the planar value \cite{roche2001, qiu2005, verzicco2006, tisserand2011, salort2014, wei2014, wagner2015}.  
In our previous study, we used roughness to break the top/bottom boundary layer symmetry, and found that a periodic upper surface with $\lambda_{\text{opt}} = 0.154$ maximized the heat transport with  $\beta = 0.359$ for a smooth lower surface in high resolution numerical simulations \cite{toppaladoddi2015_2}.   As is the case with the present geometry, when  $\lambda \ll \lambda_{\text{opt}}$ and $\lambda \gg \lambda_{\text{opt}}$, the planar results are recovered.  
For each $Ra$ we determined the maximum $Nu$ among all $\lambda$, thereby optimizing over all $\lambda$, to find $Nu_{\text{opt}} - 1 = 0.058 \times Ra^{0.334}$.

The first experimental attempt to use roughness to reach the ultimate regime at $Ra$ accessible in the laboratory was made by Roche \emph{et al.} \cite{roche2001}, who used V-shaped grooves to cover the entire interior of their cylindrical cell of $\Gamma = 0.5$. They observed a transition in $Nu(Ra)$ at $Ra \approx 2 \times 10^{12}$, and that the data beyond the point of transition could be fit with a power law with $\beta = 0.51$. A similar transition was observed at $Ra = 7 \times 10^9$ in the simulations of Stringano \emph{et al.} \cite{verzicco2006}, who used a cylindrical geometry with V-shaped grooves at the upper and lower surfaces and imposed axisymmetry on the flow. This artificial symmetry had two important effects on the flow field: (1) The production and release of the plumes from the roughness elements was in tandem, resulting in larger plumes; and (2) The plumes traversed the vertical distance without encountering a large scale circulation in the interior region. Both these effects resulted in an increase in the efficiency of the heat transfer.  As summarized by Ahlers et al., \cite[][]{ahlers2009}, it was first noted by Niemela \& Sreenivasan \cite[][]{niemela2006} that the results of Roche \emph{et al.} \cite{roche2001} can be understood as a transition between when the groove depth is less than the BL thickness to a regime where the groove depth is larger than the BL thickness.   Ahlers et al., \cite[][]{ahlers2009} state ``More work is needed to resolve this issue." Here we present results from well resolved numerical simulations of Rayleigh-B\'enard convection in a cell with rough upper and lower surfaces in two dimensions. The roughness profiles chosen are sinusoidal. By keeping the amplitude fixed and varying the wavelength of the rough surfaces, we study their effects on the heat transport.

\begin{figure}
\centering
\includegraphics[trim = 0 80 0 150, clip, width = 1\linewidth]{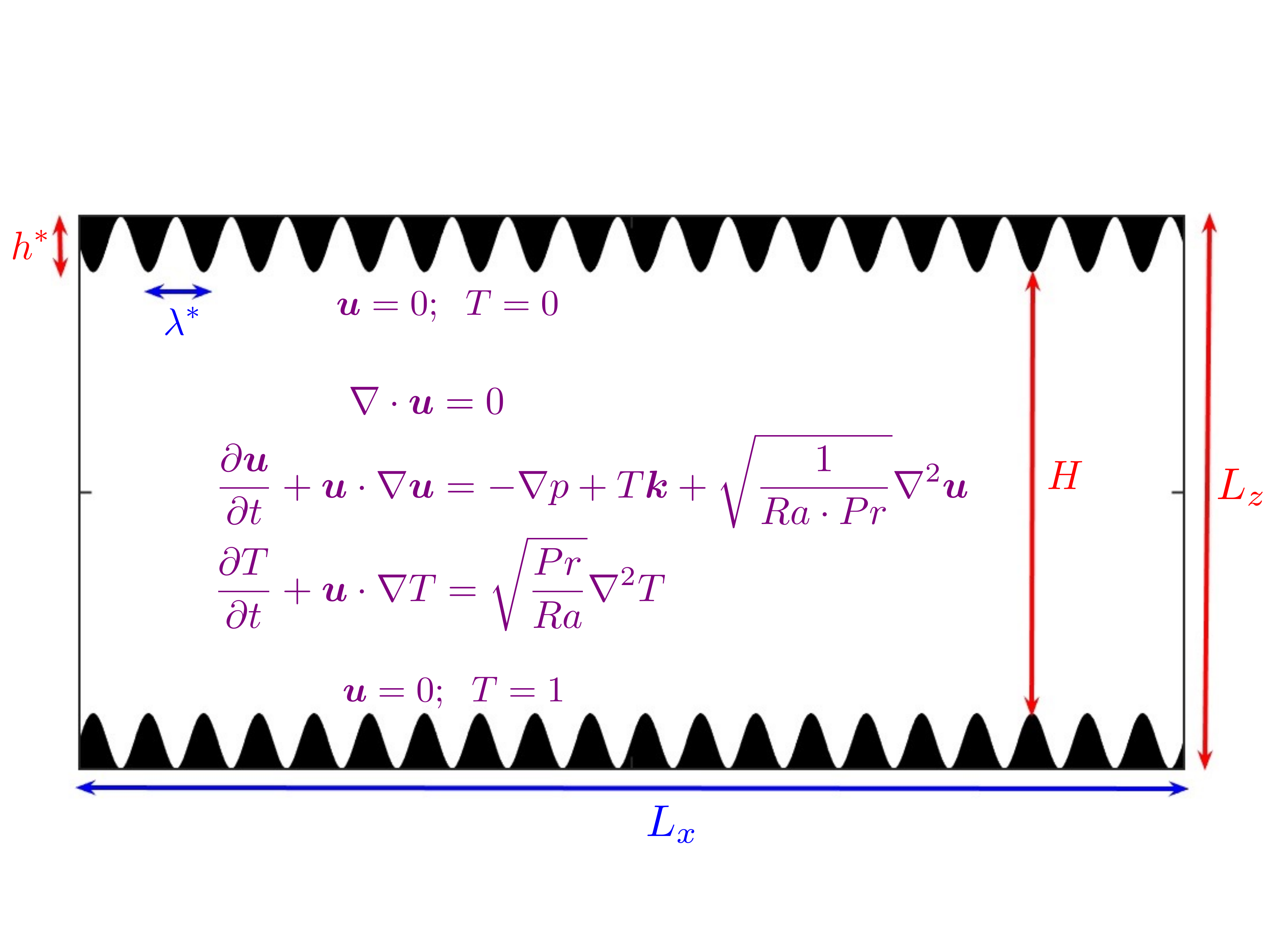}
\caption{The geometry of the rough surfaces and the equations of motion for our two-dimensional rectangular cell with $\Gamma = 2$. }
\label{fig:domain}
\end{figure}

The geometry and the dimensionless equations of motion studied here are shown in figure \ref{fig:domain}. The aspect ratio of the cell, $\Gamma \equiv L_x/L_z$, is fixed at 2. The rough surfaces have a wavelength $\lambda \equiv \lambda^*/L_z$ and an amplitude $h \equiv h^*/L_z$. The equations of motion for thermal convection are the Oberbeck-Boussinesq (O-B) equations \cite{chandra2013}, and are non-dimensionalized by choosing $H = L_z - 2h^*$ as the length scale and $U_0 = \sqrt{g\alpha \Delta T H}$ as the velocity scale. Hence, the time scale is $t_0 = H/U_0$.  Here, $\boldsymbol{u}(\boldsymbol{x},t) = \left(u(\boldsymbol{x},t), w(\boldsymbol{x},t)\right)$ is the velocity field, $T(\boldsymbol{x},t)$ is the temperature field, $\boldsymbol{k}$ is the unit vector along the vertical, and $p(\boldsymbol{x},t)$ is the pressure field. 
No-slip and Dirichlet conditions for $\boldsymbol{u}$ and $T$ are imposed on the rough surfaces, and periodic conditions are used in the horizontal.

The O-B equations were solved using the Lattice Boltzmann method with separate distributions for the momentum and temperature fields \cite{benzi1992, Massaioli:1993, chen1998, shan1997, guo2002}. Our code has been extensively tested against results from numerical simulations for a wide range of different flows, and the details of the validation can be found in \cite{toppaladoddi2015_1, toppaladoddi2015_2}.

\begin{figure}
\centering
\includegraphics[trim = 0 0 0 0, clip, width = 0.8\linewidth]{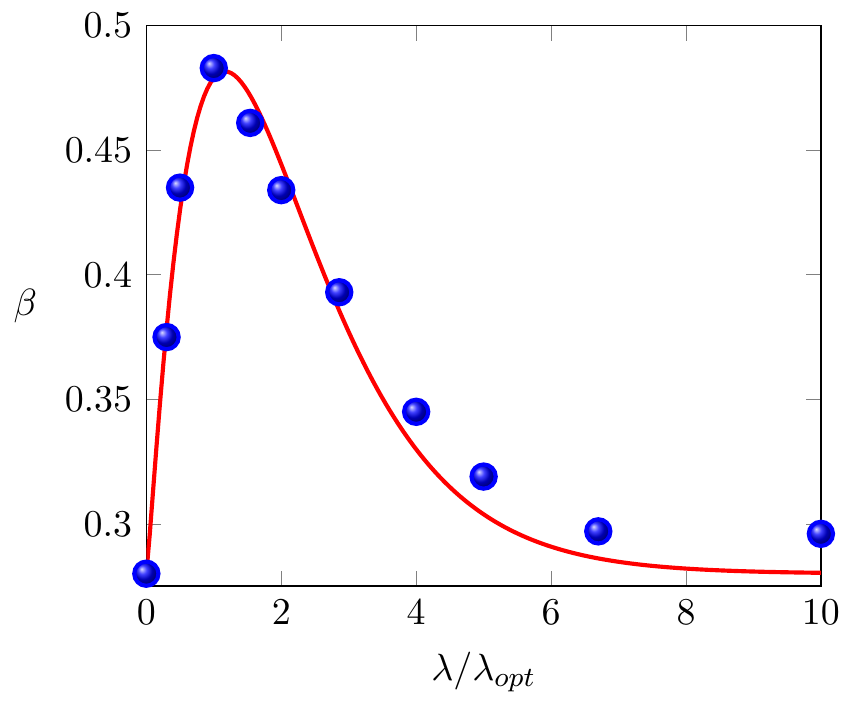}
\caption{The exponent in the scaling law $Nu-1 = A \times Ra^{\beta}$ as a function of roughness wavelength $\lambda$ (Here we used $\lambda$ = $0.03$, $0.05$, $0.1$, $0.154$, $0.2$, $0.286$, $0.4$, $0.5$, $0.67$ and $1.0$.) Data from simulations are the circles and the line is a fit using $\beta = 0.54 \, x^{1.17} \, e^{-x} + 0.28$, where $x = \lambda/\lambda_{opt}$.  At $\lambda_{\text{opt}} = 0.1$, we find a maximum $\beta_{\text{max}} = 0.483$. For $\lambda = 1$, $\beta$ is slightly larger than $0.28$ because of finite-size effects.  See also Fig. 2 of the Supplementary Material.}
\label{fig:beta}
\end{figure}

For each of ten $\lambda$'s (see Fig. \ref{fig:beta})
we simulated over the range $Ra = \left[4 \times 10^6, 3 \times 10^9\right]$. The planar wall case is $\lambda = 0$, the amplitude of the roughness is fixed at $h = 0.1$ and $Pr = 1$ for all simulations. We ran the simulations for at least $143 \, t_0$, where $t_0$ is the turnover time, and statistics were collected only after $100 \, t_0$. The Nusselt number was computed as
\begin{equation}
Nu = \frac{\left[-\kappa \frac{\partial \overline{T}}{\partial z} + \overline{w \, T}\right]_{z=z_e}}{\kappa \Delta T/H},
\label{eqn:Nu}
\end{equation}
where the overbar represents horizontal \emph{and} temporal average. We should note here that this definition of $Nu$, in general, does not reduce to unity in the static case for arbitrary roughness geometries \cite{Goluskin:2016}; however, for the sinusoidal geometries used here this choice gives $Nu \approx 1$ when $Ra = 0$. To give an example of the spatial resolutions in the simulations, for $\lambda = 1$ and $Ra = 2 \times 10^9$ the number of grid points used are $N_x = 2800$ and $N_z = 1400$. Grid independence was ascertained from simulations at $Ra = 2 \times 10^9$ for $\lambda=0.03$ and $0.2$ using two grids: (a) $N_x = 2400$, $N_z = 1200$ and (b) $N_x = 2000$, $N_z = 1000$. The difference between $Nu$ computed at $z_e = L_z/2$ for the two grids was less than $1.2 \%$. As an additional check, $Nu$ was computed at three different depths $z_e = L_z/4, L_z/2$, and $3 \, L_z/4$; and the difference between $Nu$ at any two depths was less than $0.5 \%$.  More simulation details are provided in the Supplementary Material. 

\begin{figure}
  \centering
  \subfigure{\includegraphics[scale=0.85]{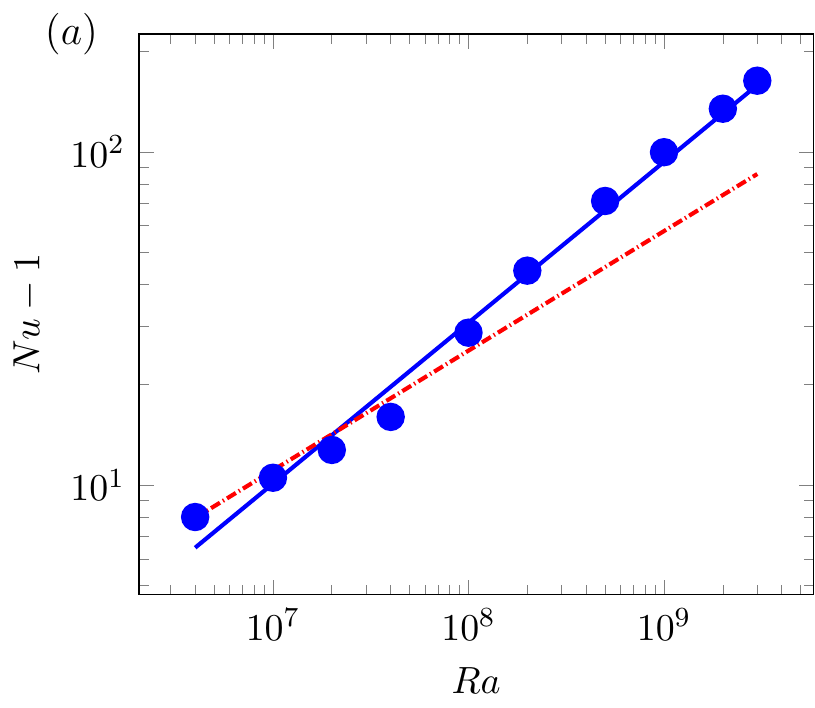}\label{fig:f1}}
  \subfigure{\includegraphics[scale=0.85]{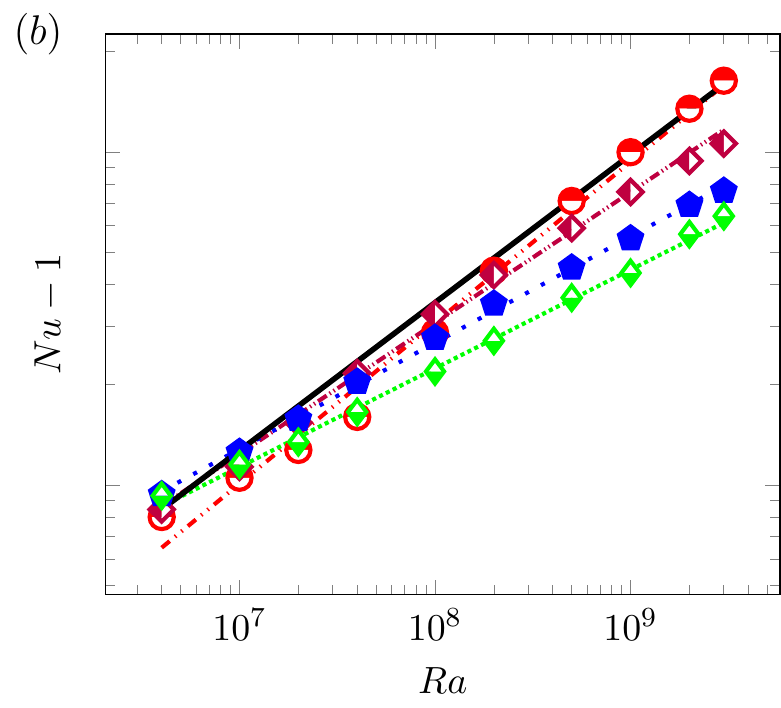}\label{fig:f2}}
  \caption{Scaling relations for different $\lambda$. (a) $Nu$-$Ra$ scaling relations for $\lambda=\lambda_{\text{opt}}=0.1$. The linear least-squares fit is $Nu - 1 = 0.0043 \times Ra^{0.482}$. The dash-dotted line is the scaling fit $Nu-1 = 0.034 \times Ra^{0.359}$ for single rough wall of $\lambda = 0.154$ \cite{toppaladoddi2015_2}. (b) The ($\lambda$, $\beta$) pairs in the order of increasing slope are ($1.0$, $0.296$), ($0.5$, $0.319$), ($0.286$, $0.393$), and ($0.1$, $0.482$). The remaining pairs (not shown in figure \ref{fig:f2}) are ($0.03$, $0.375$), ($0.05$, $0.435$), ($0.154$, $0.461$), ($0.2$, $0.434$), ($0.4$, $0.345$), and ($0.67$, $0.297$). The black line is the upper envelope is described by $Nu_{\text{opt}} - 1 = 0.01 \times Ra^{0.444}$. See also Figs. 5 and 6 in Appendix 1.}
  \label{fig:nu-ra}
\end{figure}

For each $\lambda$, we obtained $\beta$ from a linear least squares fit to the $Nu(Ra)$ simulation data. Figure \ref{fig:beta} shows $\beta$ in the scaling relation $Nu-1 = A \times Ra^{\beta}$ as a function of $\lambda$. At the optimal wavelength $\lambda_{\text{opt}} = 0.1$, $\beta$ attains a maximum value of $0.483$, which indicates that the influence of BLs on heat transport has been minimized. It is clear that in the limits $\lambda \ll \lambda_{\text{opt}}$ and $\lambda \gg \lambda_{\text{opt}}$, the planar case is approached.
The $Nu$-$Ra$ scaling relations for different $\lambda$ are shown in figure \ref{fig:nu-ra}. 
The linear least-squares fit for $\lambda_{\text{opt}} = 0.1$ giving $Nu-1 = 0.0042 \times Ra^{0.483}$ is shown in figure \ref{fig:f1}.  
The roughness elements are `submerged' inside the thermal BLs for $Ra < 10^8$ (not shown), and hence, as seen in figure \ref{fig:f2},  the values of $Nu$ for these $Ra$ are close to those for larger $\lambda$. The increase in $\beta$ for $\lambda = 0.1$ relative to other $\lambda$ is clear from figure \ref{fig:nu-ra}(b). Figure \ref{fig:f2} also shows the fit obtained for $Nu_{\text{opt}}$($Ra$), which is obtained in the following manner: for each $Ra$ we choose the maximum $Nu$ among all $\lambda$, effectively optimizing over all $\lambda$. This data is described by $Nu_{\text{opt}} - 1 = 0.01 \times Ra^{0.444}$.

\begin{figure*}
\centering
\includegraphics[trim = 0 0 0 0, clip, width = 0.8\linewidth]{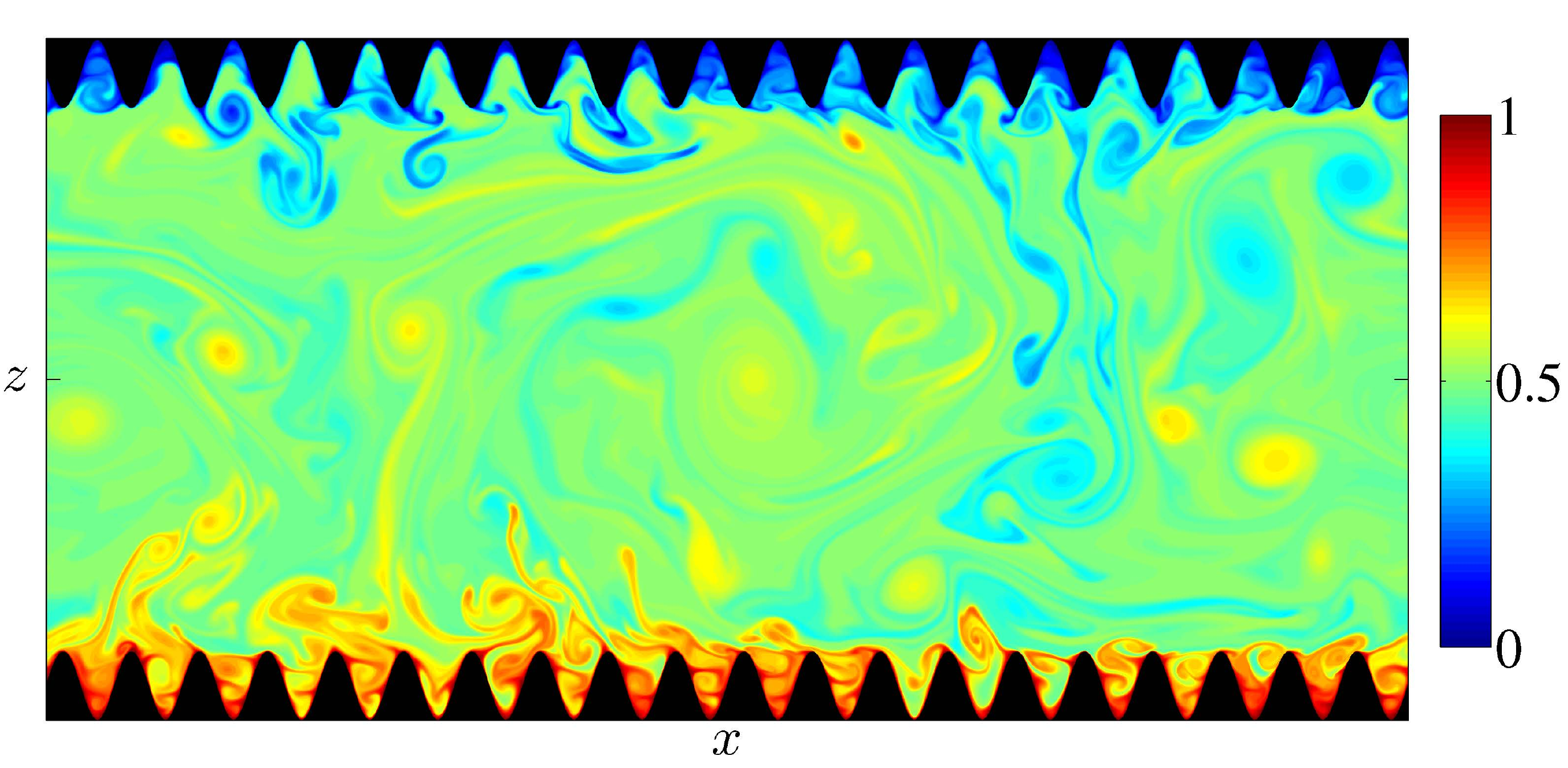}
\caption{A snapshot of the temperature field for $\lambda = 0.1$ and $Ra = 2 \times 10^9$. To see the effects of roughness, the flow field here can be
contrasted with that in the smooth case studied by Johnston \& Doering \cite{doering2009}. See also Fig. 3 of \cite{toppaladoddi2015_2} which shows the transition from the planar to the rough flow field in the case of one rough wall.}
\label{fig:temp}
\end{figure*}

The flow field for the case of $\lambda_{\text{opt}} = 0.1$ and $Ra = 2 \times 10^9$ is shown in Fig. \ref{fig:temp}, where the following features are apparent: 
\begin{enumerate}
\item Two large convection rolls in the cell interior.
\item The `unstable' BLs at the upper and lower surfaces.
\item The production of plumes from the fluid moving along the rough surfaces and their ejection from the tips of the roughness elements.
\end{enumerate}
By varying $\lambda$, we have achieved a state in which the interaction between the core flow and the BLs over the roughness elements has been enhanced. This results in an unstable state for the BLs, which then leads to the generation and ejection of plumes from the roughness tips. As noted above, in the case of a single rough wall, the maximum value of $\beta$ was found to be $\beta \approx 0.36$ \cite{toppaladoddi2015_2} but at a slightly larger $\lambda$. This highlights the role played by the second rough wall in further decreasing the role of the BLs in transporting heat. We should note here that in spite of the differences in geometry, our results have a correspondence with those of Waleffe \emph{et al.} \cite{waleffephys2015} and Sondak \emph{et al.} \cite{sondak2015} in that there is a length scale in each setting ($\lambda_{\text{opt}}$ in ours and $\alpha_{\text{opt}}$ in theirs) that optimizes heat transport. The optimization occurs through the manipulation of the coherent structures that transport heat, though in detail it is accomplished in different ways.

Our results are consistent with those of Goluskin \& Doering \cite{Goluskin:2016}, who used the background method to compute upper bounds \footnote{A detailed discussion of upper-bound studies can be found in Kerswell \cite{kerswell:1998} and Hassanzadeh \emph{et al.} \cite{hassanzadeh2014}.} on $Nu$ for R-B convection in a domain with rough upper and lower surfaces that have square-integrable gradients. They prove that $Nu \le C Ra^{1/2}$, where $C$ depends on the geometry of roughness.  Our results show that for the optimal wavelength the heat transport is $Nu-1 = 0.0042 \times Ra^{0.483}$, with the value of $C$ being four orders of magnitude larger than ours, but an exponent approaching their result. Importantly, their approach provides a key framework for exploring a range of amplitudes and wavelengths using our methodology. 
Finally, our findings demonstrate that the scaling of the ultimate regime is nearly achieved in two dimensions using rough walls. Roche \emph{et al.} \cite{roche2001} interpreted their observation of $\beta = 1/2$  as being due to a laminar to turbulent transition of the BLs.  
Here, this state is achieved by the enhanced BL--core flow interaction driven by the roughness, which generates a larger number of intense plumes.

In summary, we have studied convection in a rectangular cell of $\Gamma = 2$ with rough upper and lower surfaces. At a fixed roughness amplitude, varying the wavelength $\lambda$ results in a spectrum of exponents in the $Nu$-$Ra$ scaling relation. At $\lambda_{\text{opt}}$ the maximum exponent $\beta_{\text{max}} = 0.483$ is achieved, and in the limits $\lambda \ll \lambda_{\text{opt}}$ and $\lambda \gg \lambda_{\text{opt}}$, the planar value of $\beta$ is recovered, which may underlie why certain experiments found no effect of periodic roughness on $\beta$ \cite{du1998, du2000, ciliberto1999}.  The observation of $\beta_{\text{max}} \approx 0.5$ here has been facilitated by the use of very large amplitude roughness relative to existing studies \cite{roche2001, verzicco2006, wei2014}, indicating the promise of examining this state experimentally for more moderate values of $Ra$ than have been previously necessary.  Indeed, by varying both amplitude and wavelength over a significant range, the systematic effects of the BLs, and thus the molecular properties of the fluid, may be realized, comparing and contrasting the concept of a laminar-to-turbulent BL transition, with the enhanced forcing associated with unstable BL's triggered by the roughness as seen here.  \\ 

\begin{acknowledgements}
The authors acknowledge the support of the University of Oxford and Yale University, and the facilities and staff of the Yale University Faculty of Arts and Sciences High Performance Computing Center. 
S.T. acknowledges a NASA Graduate Research Fellowship.  J.S.W. acknowledges NASA Grant NNH13ZDA001N-CRYO, Swedish Research Council grant no. 638-2013-9243, and a Royal Society Wolfson Research Merit Award for support.
\end{acknowledgements}

\section*{Appendix 1: Optimizing heat transport over wavelength}

\begin{figure}[h!]
\centering
\includegraphics[scale=0.22]{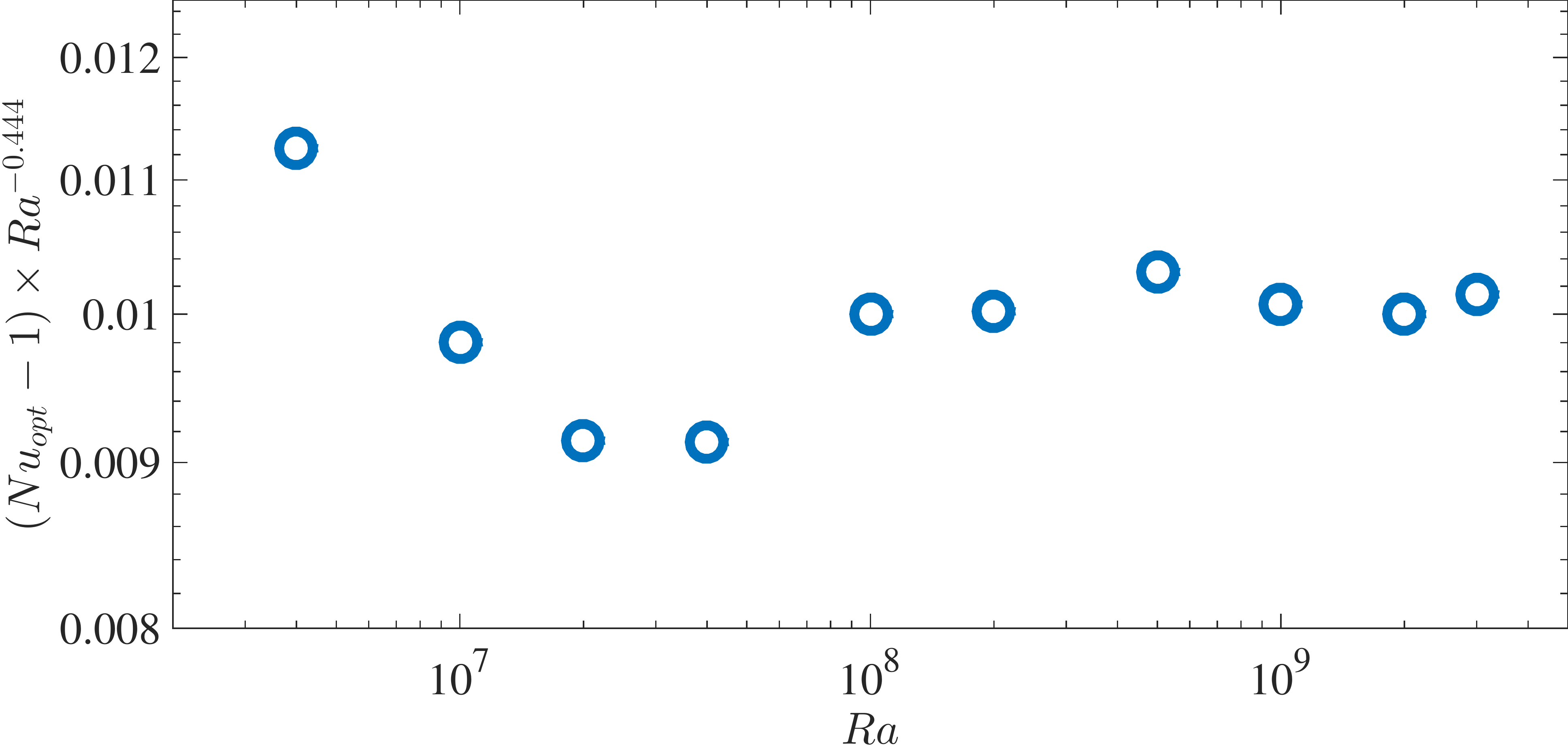} 
\caption{Compensated plot showing $\left(Nu_\text{opt}-1\right)/Ra^{0.444}$ vs. $Ra$. The prefactor is clearly $0.01$.}
\label{fig:Nu}
\end{figure}

In figure \ref{fig:Nu}, we show the compensated plot for $Nu_\text{opt}$, and 
it is apparent that the exponent for the $Nu_\text{opt}$($Ra$) scaling law is indeed $0.444$ and that 
the prefactor is $0.01$.  Whence, it provides a different means for reaching the same conclusion as described in the manuscript. 

\begin{figure}[h!]
\centering
\includegraphics[scale=0.22]{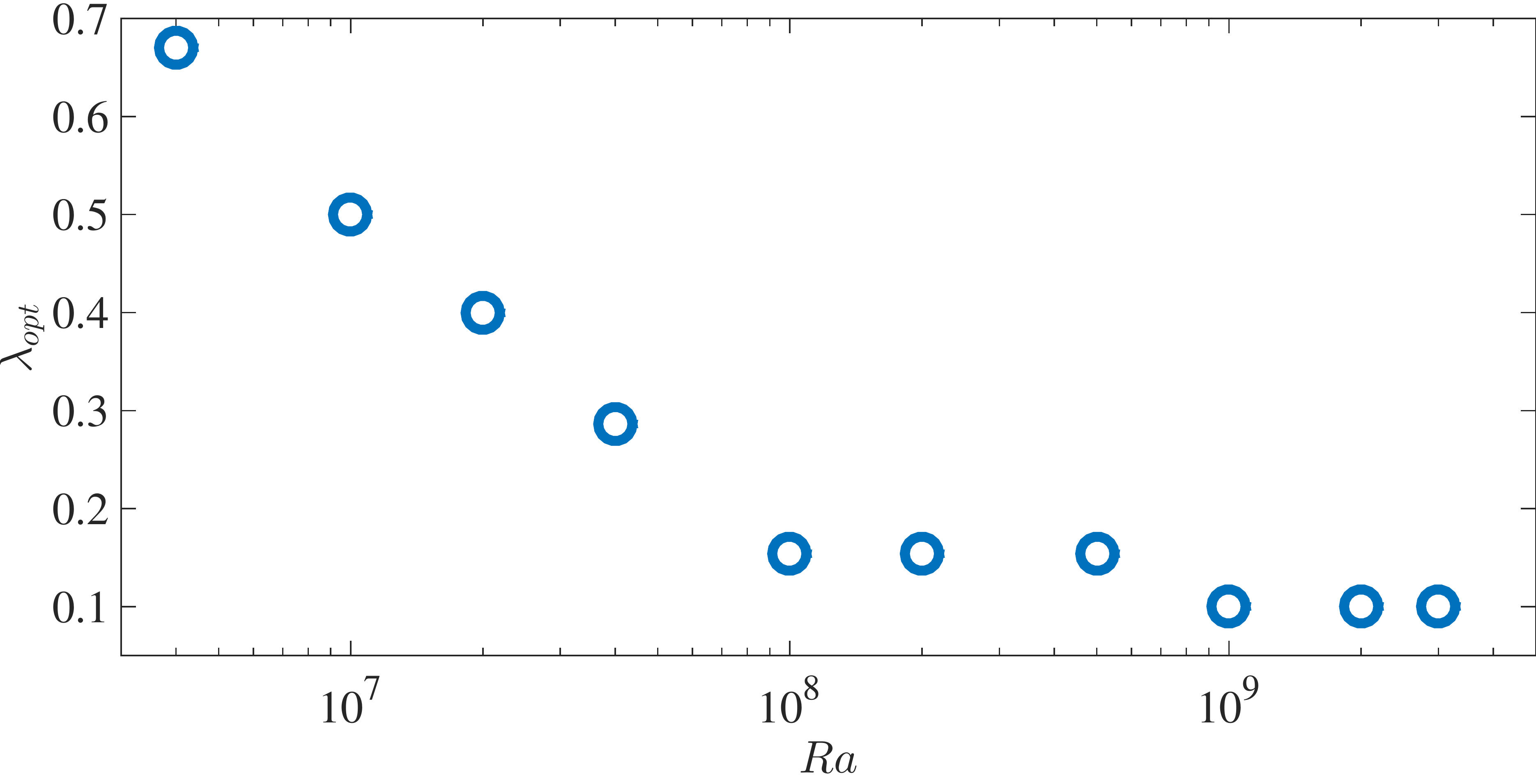} 
\caption{The variation of $\lambda_\text{opt}$ with $Ra$.}
\label{fig:lambda}
\end{figure}

Figure \ref{fig:lambda} shows the variation of $\lambda_\text{opt}$ with $Ra$. As can be seen, $\lambda_\text{opt}$ decreases from $0.67$ to $0.286$ and finally saturates to $\lambda_\text{opt} = 0.1$, implying that the wavelength for which $Nu$ is maximum for $Ra \ge 10^9$ is $0.1$. This is again consistent with figures 2 and 3b in the manuscript that show that the exponent attains a maximum value for $\lambda = 0.1$.

\section*{Appendix 2: Simulation Details}
The details of all the simulations are provided here. The roughness wavelength is $\lambda$; $Ra$ is the Rayleigh number; $N_x$ and $N_z$ are the number of grid points along the horizontal and vertical, respectively; $T_s$ is the total run time in terms of the turn-over time $t_0$; and $Nu$ is the Nusselt number.

\begin{enumerate}

\item \underline{$\lambda = 0.03$} \\~\\
    \begin{tabular}{| l | l | l | l | l |}
    \hline
    $Ra$ & $N_x$ & $N_z$ & $T_s$ & $Nu$ \\ \hline
    $4 \times 10^6$ & $1600$ & $800$ & $287.2$ & $10.50$ \\ \hline
    $10^7$ & $1600$ & $800$ & $347.0$ & $12.48$ \\ \hline
    $2 \times 10^7$ & $1600$ & $800$ & $351.8$ & $14.34$ \\ \hline
    $4 \times 10^7$ & $1600$ & $800$ & $349.3$ & $16.65$ \\ \hline   
    $10^8$ & $2000$ & $1000$ & $328.1$ & $20.92$ \\ \hline 
    $2 \times 10^8$ & $2000$ & $1000$ & $327.9$ & $26.59$ \\ \hline 
    $5 \times 10^8$ & $2000$ & $1000$ & $324.7$ & $40.18$ \\ \hline     
    $10^9$ & $2400$ & $1200$ & $230.7$ & $59.18$ \\ \hline
    $2 \times 10^9$ & $2400$ & $1200$ & $231.4$ & $90.64$ \\ \hline    
    $3 \times 10^9$ & $2800$ & $1400$ & $193.3$ & $113.44$ \\
    \hline
    \end{tabular}
    
\item \underline{$\lambda = 0.05$} \\~\\
    \begin{tabular}{| l | l | l | l | l |}
    \hline
    $Ra$ & $N_x$ & $N_z$ & $T_s$ & $Nu$ \\ \hline
    $4 \times 10^6$ & $1600$ & $800$ & $347.2$ & $9.64$ \\ \hline
    $10^7$ & $1600$ & $800$ & $345.8$ & $11.87$ \\ \hline
    $2 \times 10^7$ & $1600$ & $800$ & $348.5$ & $13.86$ \\ \hline
    $4 \times 10^7$ & $1600$ & $800$ & $347.5$ & $16.36$ \\ \hline   
    $10^8$ & $2000$ & $1000$ & $328.5$ & $21.99$ \\ \hline 
    $2 \times 10^8$ & $2000$ & $1000$ & $329.0$ & $30.68$ \\ \hline 
    $5 \times 10^8$ & $2000$ & $1000$ & $325.9$ & $55.32$ \\ \hline     
    $10^9$ & $2400$ & $1200$ & $231.2$ & $82.36$ \\ \hline
    $2 \times 10^9$ & $2400$ & $1200$ & $234.2$ & $120.26$ \\ \hline    
    $3 \times 10^9$ & $2800$ & $1400$ & $194.3$ & $127.05$ \\
\hline
    \end{tabular}
    
\item \underline{$\lambda = 0.1$} \\~\\
    \begin{tabular}{| l | l | l | l | l |}
    \hline
    $Ra$ & $N_x$ & $N_z$ & $T_s$ & $Nu$ \\ \hline
    $4 \times 10^6$ & $800$ & $400$ & $615.5$ & $9.02$ \\ \hline
    $10^7$ & $800$ & $400$ & $612.6$ & $11.52$ \\ \hline
    $2 \times 10^7$ & $800$ & $400$ & $613.0$ & $13.75$ \\ \hline
    $4 \times 10^7$ & $800$ & $400$ & $614.1$ & $17.02$ \\ \hline   
    $10^8$ & $1600$ & $800$ & $345.8$ & $29.67$ \\ \hline 
    $2 \times 10^8$ & $1600$ & $800$ & $351.6$ & $45.02$ \\ \hline 
    $5 \times 10^8$ & $2000$ & $1000$ & $315.1$ & $72.27$ \\ \hline     
    $10^9$ & $2400$ & $1200$ & $235.7$ & $100.81$ \\ \hline
    $2 \times 10^9$ & $2400$ & $1200$ & $233.8$ & $135.81$ \\ \hline    
    $3 \times 10^9$ & $2800$ & $1400$ & $201.1$ & $164.62$ \\
\hline
    \end{tabular}

\item \underline{$\lambda = 0.154$} \\~\\
    \begin{tabular}{| l | l | l | l | l |}
    \hline
    $Ra$ & $N_x$ & $N_z$ & $T_s$ & $Nu$ \\ \hline
    $4 \times 10^6$ & $800$ & $400$ & $628.9$ & $9.17$ \\ \hline
    $10^7$ & $800$ & $400$ & $630.2$ & $11.87$ \\ \hline
    $2 \times 10^7$ & $800$ & $400$ & $617.7$ & $14.25$ \\ \hline
    $4 \times 10^7$ & $800$ & $400$ & $627.1$ & $18.99$ \\ \hline   
    $10^8$ & $1600$ & $800$ & $354.1$ & $36.63$ \\ \hline 
    $2 \times 10^8$ & $1600$ & $800$ & $355.9$ & $49.60$ \\ \hline 
    $5 \times 10^8$ & $2000$ & $1000$ & $357.6$ & $76.02$ \\ \hline     
    $10^9$ & $2400$ & $1200$ & $233.1$ & $99.08$ \\ \hline
    $2 \times 10^9$ & $2400$ & $1200$ & $236.1$ & $127.52$ \\ \hline    
    $3 \times 10^9$ & $2800$ & $1400$ & $143.1$ & $139.88$ \\
\hline
    \end{tabular}

\item \underline{$\lambda = 0.2$} \\~\\
    \begin{tabular}{| l | l | l | l | l |}
    \hline
    $Ra$ & $N_x$ & $N_z$ & $T_s$ & $Nu$ \\ \hline
    $4 \times 10^6$ & $800$ & $400$ & $311.7$ & $9.27$ \\ \hline
    $10^7$ & $800$ & $400$ & $311.9$ & $12.06$ \\ \hline
    $2 \times 10^7$ & $800$ & $400$ & $312.6$ & $15.04$ \\ \hline
    $4 \times 10^7$ & $800$ & $400$ & $311.3$ & $20.22$ \\ \hline   
    $10^8$ & $1200$ & $600$ & $545.4$ & $33.38$ \\ \hline 
    $2 \times 10^8$ & $1200$ & $600$ & $549.8$ & $46.99$ \\ \hline 
    $5 \times 10^8$ & $2000$ & $1000$ & $350.2$ & $68.30$ \\ \hline     
    $10^9$ & $2400$ & $1200$ & $182.3$ & $89.55$ \\ \hline
    $2 \times 10^9$ & $2400$ & $1200$ & $235.5$ & $111.15$ \\ \hline    
    $3 \times 10^9$ & $2800$ & $1400$ & $201.5$ & $127.29$ \\
\hline
    \end{tabular}
    
\item \underline{$\lambda = 0.286$} \\~\\
    \begin{tabular}{| l | l | l | l | l |}
    \hline
    $Ra$ & $N_x$ & $N_z$ & $T_s$ & $Nu$ \\ \hline
    $4 \times 10^6$ & $800$ & $400$ & $624.1$ & $9.45$ \\ \hline
    $10^7$ & $800$ & $400$ & $614.0$ & $12.34$ \\ \hline
    $2 \times 10^7$ & $800$ & $400$ & $605.8$ & $16.40$ \\ \hline
    $4 \times 10^7$ & $800$ & $400$ & $628.1$ & $22.67$ \\ \hline   
    $10^8$ & $1600$ & $800$ & $356.3$ & $33.59$ \\ \hline 
    $2 \times 10^8$ & $1600$ & $800$ & $359.5$ & $43.71$ \\ \hline 
    $5 \times 10^8$ & $2000$ & $1000$ & $328.7$ & $60.06$ \\ \hline     
    $10^9$ & $2400$ & $1200$ & $232.2$ & $76.82$ \\ \hline
    $2 \times 10^9$ & $2400$ & $1200$ & $230.9$ & $94.98$ \\ \hline    
    $3 \times 10^9$ & $2800$ & $1400$ & $188.5$ & $106.92$ \\
\hline
    \end{tabular}

\item \underline{$\lambda = 0.4$} \\~\\
    \begin{tabular}{| l | l | l | l | l |}
    \hline
    $Ra$ & $N_x$ & $N_z$ & $T_s$ & $Nu$ \\ \hline
    $4 \times 10^6$ & $800$ & $400$ & $621.3$ & $9.81$ \\ \hline
    $10^7$ & $800$ & $400$ & $611.2$ & $13.28$ \\ \hline
    $2 \times 10^7$ & $800$ & $400$ & $621.1$ & $16.94$ \\ \hline
    $4 \times 10^7$ & $800$ & $400$ & $623.1$ & $21.94$ \\ \hline   
    $10^8$ & $1600$ & $800$ & $364.4$ & $30.77$ \\ \hline 
    $2 \times 10^8$ & $1600$ & $800$ & $358.0$ & $38.59$ \\ \hline 
    $5 \times 10^8$ & $2000$ & $1000$ & $325.7$ & $51.05$ \\ \hline     
    $10^9$ & $2400$ & $1200$ & $218.9$ & $63.94$ \\ \hline
    $2 \times 10^9$ & $2400$ & $1200$ & $222.2$ & $78.21$ \\ \hline    
    $3 \times 10^9$ & $2800$ & $1400$ & $179.5$ & $85.82$ \\
\hline
    \end{tabular}

\item \underline{$\lambda = 0.5$} \\~\\
    \begin{tabular}{| l | l | l | l | l |}
    \hline
    $Ra$ & $N_x$ & $N_z$ & $T_s$ & $Nu$ \\ \hline
    $4 \times 10^6$ & $800$ & $400$ & $620.4$ & $10.38$ \\ \hline
    $10^7$ & $800$ & $400$ & $625.6$ & $13.57$ \\ \hline
    $2 \times 10^7$ & $800$ & $400$ & $626.1$ & $16.80$ \\ \hline
    $4 \times 10^7$ & $800$ & $400$ & $612.2$ & $21.40$ \\ \hline   
    $10^8$ & $1600$ & $800$ & $361.5$ & $28.65$ \\ \hline 
    $2 \times 10^8$ & $1600$ & $800$ & $357.9$ & $36.00$ \\ \hline 
    $5 \times 10^8$ & $2000$ & $1000$ & $325.8$ & $45.93$ \\ \hline     
    $10^9$ & $2400$ & $1200$ & $228.4$ & $55.98$ \\ \hline
    $2 \times 10^9$ & $2400$ & $1200$ & $215.5$ & $70.26$ \\ \hline    
    $3 \times 10^9$ & $2800$ & $1400$ & $185.5$ & $77.39$ \\
\hline
    \end{tabular}

\item \underline{$\lambda = 0.67$} \\~\\
    \begin{tabular}{| l | l | l | l | l |}
    \hline
    $Ra$ & $N_x$ & $N_z$ & $T_s$ & $Nu$ \\ \hline
    $4 \times 10^6$ & $800$ & $400$ & $630.3$ & $10.60$ \\ \hline
    $10^7$ & $800$ & $400$ & $628.0$ & $13.29$ \\ \hline
    $2 \times 10^7$ & $800$ & $400$ & $606.3$ & $16.34$ \\ \hline
    $4 \times 10^7$ & $800$ & $400$ & $620.3$ & $19.99$ \\ \hline   
    $10^8$ & $1600$ & $800$ & $357.5$ & $26.20$ \\ \hline 
    $2 \times 10^8$ & $1600$ & $800$ & $363.0$ & $30.32$ \\ \hline 
    $5 \times 10^8$ & $2000$ & $1000$ & $304.4$ & $40.36$ \\ \hline     
    $10^9$ & $2400$ & $1200$ & $217.7$ & $50.01$ \\ \hline
    $2 \times 10^9$ & $2400$ & $1200$ & $213.1$ & $62.21$ \\ \hline    
    $3 \times 10^9$ & $2400$ & $1200$ & $213.2$ & $68.66$ \\
\hline
    \end{tabular}
    
\item \underline{$\lambda = 1.0$} \\~\\
    \begin{tabular}{| l | l | l | l | l |}
    \hline
    $Ra$ & $N_x$ & $N_z$ & $T_s$ & $Nu$ \\ \hline
    $4 \times 10^6$ & $800$ & $400$ & $508.1$ & $10.29$ \\ \hline
    $10^7$ & $800$ & $400$ & $506.8$ & $12.54$ \\ \hline
    $2 \times 10^7$ & $800$ & $400$ & $509.9$ & $14.46$ \\ \hline
    $4 \times 10^7$ & $800$ & $400$ & $502.8$ & $17.58$ \\ \hline   
    $10^8$ & $1600$ & $800$ & $351.6$ & $22.93$ \\ \hline 
    $2 \times 10^8$ & $1600$ & $800$ & $368.9$ & $28.10$ \\ \hline 
    $5 \times 10^8$ & $2000$ & $1000$ & $297.6$ & $37.51$ \\ \hline     
    $10^9$ & $2400$ & $1200$ & $207.6$ & $44.34$ \\ \hline
    $2 \times 10^9$ & $2400$ & $1200$ & $203.3$ & $57.66$ \\ \hline    
    $3 \times 10^9$ & $2800$ & $1400$ & $173.4$ & $65.18$ \\
\hline
    \end{tabular}
\end{enumerate}

\section*{Appendix 3: Grid Independence Tests}
The following tests were performed to ascertain the grid independence of the results:
\begin{enumerate}

\item \underline{$\lambda = 0.1$ and $Ra = 2 \times 10^8$} \\
Two grids were used: (1) $N_x = 2400$ and $N_z = 1200$ and (2) $N_x = 1600$ and $N_z = 800$. The difference in $Nu$ from these two runs was $0.3 \%$.

\item \underline{$\lambda = 0.2$ and $Ra = 10^9$} \\
Two grids were used: (1) $N_x = 2400$, $N_z = 1200$ and (2) $N_x = 1600$, $N_z = 800$. The difference in $Nu$ between these two runs was $3.7\%$.

\item \underline{$\lambda = 0.03$ and $Ra = 2 \times 10^9$} \\
Two grids were used: (1) $N_x = 2400$ and $N_z = 1200$ and (2) $N_x = 2000$ and $N_z = 1000$. The difference in $Nu$ from these two runs was $0.9 \%$.

\item \underline{$\lambda = 0.2$ and $Ra = 2 \times 10^9$} \\
Two grids were used: (1) $N_x = 2400$, $N_z = 1200$ and (2) $N_x = 2000$, $N_z = 1000$. The difference in $Nu$ between these two runs was $0.1\%$.

\end{enumerate}

We note here that the smaller grid used in test run 2 for $Ra = 10^9$ was mainly to check the robustness of the code. Such a large $Ra$ in general requires more number of grid points to resolve the flow in the roughness region. The difference in $Nu$ of $3.7\%$ with the higher resolution run demonstrates that the numerical method employed is adequately robust.

Additionally, the code has been thoroughly validated against results from spectral codes for both Rayleigh-B\'enard convection and transitional flows in two-dimensional channels \cite{toppaladoddi2015_2, toppaladoddi2015_1}.


%

\end{document}